\documentclass[conference]{IEEEtran}
\IEEEoverridecommandlockouts
\usepackage{cite}
\usepackage{amsmath,amssymb,amsfonts}
\usepackage{algorithmic}
\usepackage{graphicx}
\usepackage{textcomp}
\usepackage{xcolor}
\usepackage{subcaption}
\usepackage{caption}
\def\BibTeX{{\rm B\kern-.05em{\sc i\kern-.025em b}\kern-.08em
    T\kern-.1667em\lower.7ex\hbox{E}\kern-.125emX}}

\usepackage{comment}

\usepackage{multirow}
\usepackage{tabularx}

\usepackage[english]{babel}  
\usepackage{microtype}  
\usepackage[colorlinks=true, linkcolor=black, citecolor=black, urlcolor=black]{hyperref}
\hypersetup{breaklinks=true}
\usepackage{soul}
\usepackage{enumitem}
\usepackage[subtle]{savetrees}
\setlist[itemize]{itemsep=0pt, topsep=2pt, leftmargin=10pt}

\usepackage{tikz} \newcommand*\circled[1]{\tikz[baseline=(char.base)]{\node[shape=circle,fill,inner sep=1pt] (char) {\textcolor{white}{#1}};}}
\usepackage{arydshln}

\begin{document}

\title{Efficient Fine-Grained GPU Performance Modeling for Distributed Deep Learning of LLM\\
\thanks{This work was supported in part by the NSF research grant \#2320952, \#2117439, \#2112606, and \#2117439.}
}

\author{
\IEEEauthorblockN{Biyao Zhang\textsuperscript{1}, Mingkai Zheng\textsuperscript{2}, Debargha Ganguly\textsuperscript{1}, Xuecen Zhang\textsuperscript{1}, Vikash Singh\textsuperscript{1}}
\IEEEauthorblockA{Vipin Chaudhary\textsuperscript{1}, Zhao Zhang\textsuperscript{2}}
\\
\IEEEauthorblockA{\textsuperscript{1}Case Western Reserve University, Cleveland, OH, USA \textsuperscript{2}Rutgers University, Brunswick, NJ, USA\\
\{biyao.zhang, debargha, xuecen.zhang, vikash, vipin\}@case.edu, \{mingkai.zheng, zhao.zhang\}@rutgers.edu}\\
}

\maketitle

\begin{abstract}
Training Large Language Models(LLMs) is one of the most compute-intensive tasks in high-performance computing. Predicting end-to-end training time for multi-billion parameter models distributed across hundreds of GPUs remains challenging due to complex interactions between transformer components, parallelism strategies(data, model, pipeline, tensor), and multi-tier communication. Learned models require costly sampling, while analytical models often struggle with real-world network and hardware complexities.
We address this by decomposing LLMs into core computational primitives and modeling them with: (1) operator-level decomposition for fine-grained analysis; (2) lightweight sampling based hardware-aware prediction models for key operations; (3) an end-to-end prediction system integrating these components across complex parallelization strategies.
Crucially, our methodology has been validated on two large-scale HPC systems. Our framework achieves low average prediction errors-4.98\% on Perlmutter(A100) and 9.38\% on Vista(GH200)-for models up to 20B parameters across 128 GPUs. Importantly, it runs entirely on CPUs, enabling rapid iteration over hardware configurations and training strategies without costly on-cluster experimentation.

\end{abstract}




\section{Introduction}

Large Language Models (LLMs) have evolved into one of the most computationally intensive workloads in high-performance computing (HPC), with resource requirements increasing 10-100x per model generation. Foundation models such as GPT-4 and LLaMA 3 require on the order of 10$^{25}$ FLOPs, requiring months of training on tens of thousands of high-end GPUs and incurring costs in the hundreds of millions of dollars~\cite{brown2020language, hoffmann2022training}. This unprecedented scaling, predicted by various scaling laws~\cite{kaplan2020scaling, hoffmann2022training, wu2024inference}, has transformed LLM training into a critical HPC scheduling and resource allocation challenge. System operators and researchers must make high-stakes decisions about computational investments months before training begins, with each miscalculation potentially costing millions in wasted resources or delayed research timelines. \emph{Effective planning and optimization of these large-scale training runs hinges on accurately modeling and accurately predicting performance across complex supercomputing environments before committing substaintial resources.}

Accurately predicting training time for large-scale LLMs distributed across HPC systems using complex parallelism is uniquely challenging, surpassing prior modeling focused on memory estimation~\cite{rajbhandari2020zero, rajbhandari2021zero}, smaller networks~\cite{justus2018predicting, li2023path}, or specific distributed strategy optimization~\cite{yun2023fast}. A significant gap remains in providing accurate, empirically validated end-to-end time predictions for these multi-billion parameter pre-training jobs on real-world supercomputers. Models developed for simpler scenarios like single-GPU fine-tuning lack the required distributed complexity. Furthermore, purely analytical approaches, while insightful, often struggle to capture real-world performance impacts from opaque hardware optimizations, system jitter, and critically, the complex, stochastic behavior of communication—including collectives and pipelined transfers—across actual, potentially congested HPC network fabrics. Consequently, reliably forecasting performance for these demanding workloads before resource commitment remains a critical unmet need.

These challenges have led to the adoption of sampling-based learned performance models~\cite{kaufman2021learned, gao2020estimating, geoffrey2021habitat, zhu2020daydream}, which run abbreviated training on target hardware to estimate full-scale performance. However, this approach incurs prohibitive costs: sampling just 60 seconds of training for a 20B parameter model on 128 A100 GPUs consumes two node-hours, with multi-configuration exploration potentially consuming 5-10\% of the total training budget~\cite{kaufman2021learned}—creating additional strain on already GPU-limited HPC queues. Moreover, LLM performance modeling faces three key challenges: (1) heterogeneous GPU architectures exhibiting nonlinear scaling and containing opaque, vendor-specific optimizations; (2) diverse computational patterns within transformers spanning GEMM operations and bandwidth-sensitive normalization layers; and (3) complex, often overlapping interactions between computation and communication arising from sophisticated parallelism strategies (data, tensor, pipeline) deployed across multi-tier interconnects. Accurately modeling communication, especially overlapped pipeline stages and large-scale collectives, remains a major hurdle for purely analytical approaches in predicting real-world execution times. These factors create a significant gap between abstract computational graphs or simplified models and actual performance, limiting the utility of conventional black-box or overly-simplistic analytical approaches.

We propose that systematic decomposition can accurately model \textit{large-scale pre-training} performance without costly end-to-end sampling. By profiling core computational primitives across hardware and integrating their predictions, our method achieves accurate, hardware-aware forecasts with minimal overhead—significantly more efficient than existing approaches.
Our detailed contributions include:

\vspace{1ex}

\noindent \circled{1} We introduce \ul{operator-level decomposition} to decompose transformer-based LLM architectures into fundamental operators, enabling fine-grained performance analysis while maintaining analytical tractability. This approach allows for precisely modeling hardware-specific behaviors and non-linear performance patterns across diverse model architectures.

\vspace{1ex}

\noindent \circled{2} We develop \ul{lightweight sampling based hardware-aware performance prediction models}, which are specialized regression models tailored to different operation types (compute-bound matrix multiplications, memory-bound normalizations, and network-dependent communications), capturing the unique performance characteristics of each component. Our data collection methodology balances accuracy, computational feasibility, and parameter space coverage to enable robust predictions.

\vspace{1ex}

\noindent \circled{3} We implement an \ul{end-to-end training time prediction system} integrating component-level predictions to accurately model complete LLM training workflows across complex parallelization strategies (data, model, and pipeline). Our framework demonstrates high accuracy with average prediction errors of 4.98\% on Perlmutter (up to 128 NVIDIA A100-SXM4 GPUs) and 9.38\% on Vista (up to 128 NVIDIA GH200 GPUs) for state-of-the-art models, including GPT-20B, LLaMA-13B, and Llemma-7B, validated directly on these target HPC systems.

Our methodology helps optimize large-scale AI workloads by predicting performance and identifying hardware bottlenecks across parallelization strategies. It improves time-to-solution and system throughput, making it especially valuable for pre-exascale and exascale systems where LLMs dominate.






\section{Foundations and Problem Statement}
This section lays the foundation for our performance modeling by covering GPU architectures, execution models, transformer computation patterns, and distributed training strategies, leading to the key challenges addressed in this paper.
We adopt GPT-NeoX~\cite{black2022gptneox20bopensourceautoregressivelanguage} which integrated DeepSpeed and Megatron-LM as the framework for implementation of 3D parallelism.

\subsection{Distributed Training Systems} 
\label{ss:background-distributed-training}

\vspace{1ex}
\noindent \textbf{Modern GPUs:} Modern NVIDIA GPUs feature CUDA cores for general computation and Tensor Cores for matrix operations within Streaming Multiprocessors (SMs) that share a hierarchical memory system. Each SM has private L1 cache and shared memory, while all SMs access L2 cache and global memory, creating operation-dependent bottlenecks.
GPU execution relies on warp-level parallelism under SIMT, with A100 and H100 optimizing scheduling through multiple warp slots per SM to hide latency. Performance depends on thread configuration, memory access, and hardware tuning.
Matrix performance varies by GPU. A100 supports TF32 and BF16, while H100 adds FP8 and doubles CUDA cores per sub-partition. Vendor auto-tuning in cuBLAS and cuDNN causes discontinuous scaling, requiring empirical profiling over analytical models.

\vspace{1ex}
\noindent \textbf{Transformer Architecture and Computational Patterns:} 
The transformer architecture stands as the foundation of modern language models.
Each transformer block contains several key components with different execution characteristics:

1) \textit{Multi-head attention (MHA)} generates diverse matrix multiplication patterns with dimensions determined by batch size, sequence length, hidden dimension, and head count. These operations exhibit complex performance scaling due to memory access patterns and auto-tuning optimizations.

2) \textit{Multi-layer perceptron (MLP)} typically expand the hidden dimension by 4×, creating large matrix operations with different performance characteristics than attention layers.

3) \textit{Normalization layers} (LayerNorm, RMSNorm) perform fine-grained elementwise operations that are often memory-bandwidth limited rather than compute-bound.

These components create a heterogeneous computational graph where different operations can alternate between compute-bounded and memory-bounded depending on computing hardware, batch size, and model dimensions. This heterogeneity necessitates operator-specific performance modeling approaches. \emph{We specifically model the Parallel Self-Attention and Parallel MLP components from GPT-NeoX~\cite{black2022gptneox20bopensourceautoregressivelanguage}, excluding communication mechanisms. }

\vspace{1ex}
\noindent \textbf{Distributed Training Strategies:} 
Scaling language model training across GPU clusters requires coordinated parallelization strategies to overcome memory limitations and optimize computational throughput. Modern distributed training frameworks implement three primary parallelism dimensions:

\textit{Data Parallelism (DP)} replicates the model across devices, with each replica processing different data batches. This approach requires synchronizing gradients through all-reduce operations, with communication overhead scaling linearly with model size and device count. Optimization techniques like DeepSpeed ZeRO~\cite{10.1145/3394486.3406703} improve memory efficiency by partitioning optimizer states and gradients across devices.

\textit{Model Parallelism (MP)} distributes model parameters across devices, enabling the training of models that exceed single-GPU memory capacity. A subset named Tensor-parallel, used in frameworks like Megatron-LM~\cite{shoeybi2020megatronlmtrainingmultibillionparameter}, partitions individual layers by splitting matrix dimensions, requiring all-reduce operations to synchronize partial results. This creates intricate dependencies between computation and communication that vary by layer type and position.

\textit{Pipeline Parallelism (PP)} segments model layers across devices, enabling concurrent processing of different micro-batches through sequential pipeline stages. This approach introduces complex scheduling considerations, including pipeline bubble minimization and load balancing across heterogeneous stages. The efficiency of pipeline parallelism depends critically on micro-batch count, stage distribution, and inter-stage communication costs.

These parallelization strategies can be combined in "3D parallelism" frameworks that simultaneously optimize along all three dimensions. However, this introduces complex interactions between computation and communication patterns that significantly impact training performance.

\begin{figure*}[t]
    \centering
    \includegraphics[width=1\linewidth]{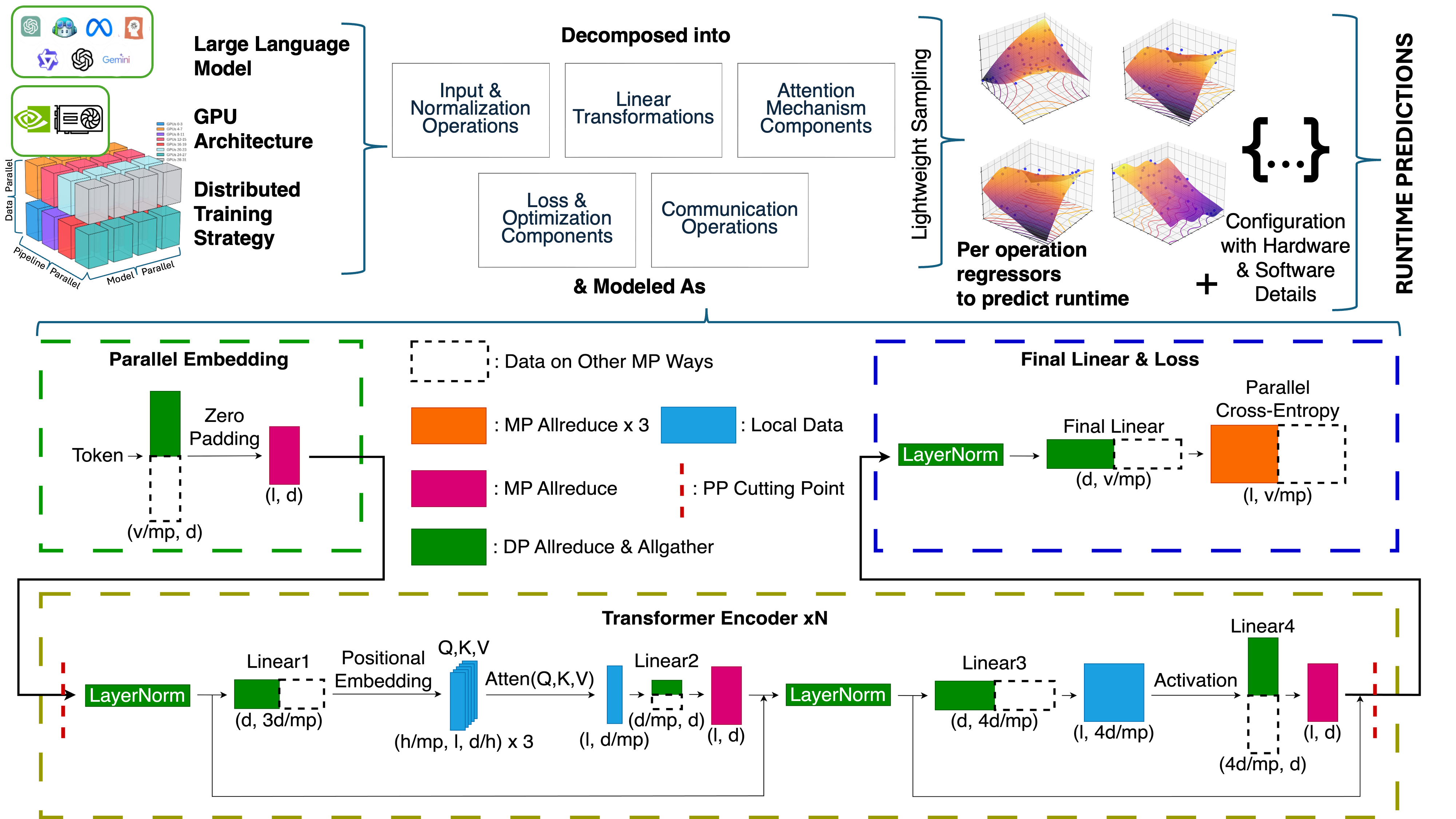}
    \caption{Performance modeling workflow for distributed LLM training: The approach decomposes 3D parallelism into fundamental operations across three modules (embedding, transformer encoder, loss), collects per-operation performance data, builds specialized regressors, and integrates predictions for end-to-end modeling. Input/output shape annotations show data flow and memory usage, while color coding distinguishes data segments for collective operations synchronizing intermediate data and gradients. Abbreviations: PP – Pipeline Parallelism, MP – Model Parallelism, DP – Data Parallelism.}
    \vspace{-2ex}
    \label{fig:workload}
\end{figure*}

\subsection{Problem Statement}
\label{ss:problem-statement}

As demonstrated in Section~\ref{ss:background-distributed-training}, performance modeling of distributed deep learning presents several interconnected challenges that arise from the complexity of modern GPU architectures, the computational patterns of transformer-based models, and the intricacies of multi-tier parallelism strategies. Accurately predicting training performance requires addressing the following three fundamental challenges:

\vspace{1ex}
\noindent \textbf{\textit{Challenge 1: How can we model performance across heterogeneous GPU architectures?}}

Modern GPU clusters vary widely in architecture. Differences between NVIDIA’s A100 and H100—such as tensor core throughput, memory bandwidth, cache, and interconnects—cause nonlinear performance scaling. GPU libraries like cuBLAS and NCCL add further variability through opaque, hardware-specific optimizations. As a result, analytical models often fail to generalize, making efficient empirical modeling essential.

Mixed precision training adds complexity: formats like FP32, FP16, and BF16 impact compute and memory differently. Quantization introduces conversion overhead but may reduce memory bandwidth needs, complicating cross-platform performance prediction.

\vspace{1ex}
\noindent \textbf{\textit{Challenge 2: How can we model the diverse computational patterns in transformer architectures?}}

Transformer-based models, especially LLMs, introduce significant performance modeling challenges due to their complex layer structure. Components like multi-head attention (MHA), feedforward, normalization, and activation layers each exhibit unique compute and memory patterns. MHA layers produce diverse General Matrix Multiply (GEMM) shaped by input size and model config, while normalization and activations rely on memory-bound, fine-grained operations. These varied bottlenecks hinder the effectiveness of a unified performance model.
Matrix multiplications in transformers exhibit discontinuous performance due to GPU auto-tuning and kernel switching based on input shapes, leading to step-like performance curves. Normalization and activation layers, being memory-bound, show complex scaling tied to batch size and cache behavior, differing from compute-bound operations and requiring distinct modeling.
Memory hierarchy optimization techniques such as gradient checkpointing, activation recomputation, and offloading introduce complex memory access patterns that significantly impact performance. These techniques create trade-offs between computation redundancy and memory consumption that require specialized modeling approaches to capture accurately.

\vspace{1ex}
\noindent \textbf{\textit{Challenge 3: How can we model the interplay of parallelism strategies and communication overheads?}}

Scaling deep learning workloads across multiple GPUs and nodes requires a combination of data parallelism (DP), model parallelism (MP), and pipeline parallelism (PP). Each strategy introduces distinct synchronization and communication costs. Specifically, DP may suffer from all-reduce synchronization bottlenecks that depend on GPU count and network topology; MP requires inter-GPU communication for partial sum reductions, which can incur latency and bandwidth overheads; and PP introduces inter-stage dependencies that can lead to pipeline bubbles, degrading efficiency. Communication operations (all-reduce, all-gather) introduce additional complexity due to network topology, collective algorithm choice, and runtime conditions, including network congestion and concurrent computation. These operations are susceptible to system configuration and load, making them challenging to model consistently.


Moreover, communication efficiency is impacted by \textit{multi-tier interconnect heterogeneity}, where NVLink, PCIe, and Infiniband/Ethernet introduce varying bandwidth and latency constraints. These complex dependencies make traditional scaling laws insufficient for accurately predicting distributed training performance.

The confluence of these factors creates a significant gap between the abstract computational graph of an LLM and its realized performance on a multi-tier HPC infrastructure. Conventional black-box modeling approaches—such as fitting iteration time to GPU count or hidden dimension size—fail to capture the intricate interplay between computational, memory, and communication bottlenecks. Performance does not scale in isolation; instead, it emerges from the nonlinear interactions of compute-bound, memory-bound, and communication-bound operations across heterogeneous architectures. 

These challenges motivate our operator-level decomposition, which models each component independently and integrates them into a unified performance model, enabling accurate predictions across diverse hardware and architectures with minimal profiling.

\section{Methodology}

We use a bottom-up approach to performance modeling(as shown in Figure~\ref{fig:workload}), starting with operator-level profiling of core computations like matrix multiplications, attention, and communication. Per-operator regressors capture compute, memory, and communication behavior across hardware. These models are hierarchically aggregated to predict end-to-end performance. 

\subsection{Performance Data Collection}
\label{ss:performance-data-collection}

We propose a \textit{three-pronged strategy}—micro-benchmarking, parameter exploration, and interpolation—to build a robust LLM performance dataset that captures nonlinear, hardware-specific behaviors beyond the reach of theoretical models.


\vspace{1ex}
\noindent \textbf{Micro-benchmark Design:} Operator-level performance is measured using micro-benchmarks that reflect real LLM workloads. For the training framework, GPT-NeoX, All operators-including both computation and communication operators-are executed in isolation without any kernel level overlap, ensuring that each kernel function can fully utilize all GPU resources for minimum execution time and the influence of prior kernels. These operators are profiled with varying input features and key hyperparameters recorded alongside latency, enabling precise analysis without interference from the framework level.


\vspace{1ex}
\noindent \textbf{Profiling and Measuring Infrastructure:} We utilize the PyTorch profiler with 1$\mu$s recording resolution and CUDA event recording. Our custom parser identifies target operators by searching for specific dictionary items in the profiler output. It calculates total GPU runtime as the difference between maximum end time and minimum start time of all associated kernels, capturing complete execution time including memory transfers. The profiling begins with a 10-iteration warmup phase using an input configuration designed to saturate the GPU’s capability, followed by an additional 10 iterations to ensure steady-state execution. To mitigate the influence of outliers, the mean of sorted median 5 samples are selected as final measurement.



\vspace{1ex}
\noindent \textbf{Operator and Communication Sampling:} We isolated operators via source code extraction and profiler-based sampling to capture framework-specific optimizations. For distributed operators, key communication patterns were benchmarked across layouts to reflect topology effects, with sampling adapted to hardware while keeping processor counts comparable.

\subsection{Regressors for Fundamental Operations}
\vspace{1ex}
Tree-based regressors interpolate and extrapolate performance from collected data, capturing execution discontinuities from auto-tuning, memory hierarchy, and collective algorithms. Validation against real training runs ensures alignment with actual GPU performance across architectures and parallelism strategies.
We use RandomForest and XGBoost for operator-level performance modeling as they capture non-linear GPU performance patterns influenced by hardware thresholds. Their hierarchical structure efficiently models piecewise relationships, offering fast inference and interpretability over neural networks. Their proven robustness with tabular data helps reveal how dimensions, batch size, and concurrency affect runtime, aiding validation and optimization. To avoid overfitting, we select the regressor and its hyperparameters for each operator based on the principle of minimizing validation error, using 80\% of the data for training and 20\% for validation. Once selected, the final regressor is built on the entire dataset.

\subsection{Workload Representation}

Distributed LLM training performance depends on compute, memory, and communication patterns, which vary by hardware and model. We design a structured workload representation to capture operator behavior, parallelism, and system constraints for accurate modeling.

\vspace{1ex}
\noindent \textbf{Operator-Level Representation:}
Each transformer layer is characterized by batch size ($b$), which determines the number of samples processed per iteration; sequence length ($l$), affecting attention computation cost; hidden dimension ($d$), defining the per-token feature size and influencing matrix multiplications; attention heads ($h$), controlling query, key, and value projection splits in multi-head self-attention; and model parallelism degree ($|mp|$), governing weight partitioning across GPUs, impacting memory distribution and communication overhead.

Beyond model-specific parameters, workload representation incorporates distributed training characteristics. $|entries|$, $|nodes|$, and $|GPUs/nodes|$ denote the total number of data entries in operations (like all-reduce), the number of compute nodes, and GPUs per node, respectively, define interconnect topology and parallelism efficiency. The number of transformer encoder layers, $|encoders|$, governs gradient aggregation and optimizer updates. Lastly, $dim$ represents model parameter dimensionality, affecting memory footprint and weight update costs.

For example, linear layers perform affine transformations essential for feature projection and dimensionality changes in transformer architectures. The Linear1 layer applies a weight matrix projection, transforming an input tensor of shape $[bl, d]$ (where $bl$ represents the combined batch and sequence dimensions, and $d$ is the hidden size) into an output of shape $[bl, 3d/|mp|]$. This transformation expands the hidden dimension threefold, corresponding to the query, key, and value (QKV) projections used in multi-head self-attention (MHSA). Each of these projected representations is subsequently split into multiple attention heads and processed separately to capture diverse relational patterns between tokens. Table~\ref{tab:Feature Selection for Regressor} details the regressor input vector, mapping each operator type to its corresponding workload representation.

\begin{table}[htbp]
\small
\caption{Feature selection methodology for performance prediction regressors. The table details the workload representation vectors used for each operator type, incorporating batch size ($b$), sequence length ($l$), hidden dimension ($d$), attention heads ($h$), model parallel degree ($|mp|$), and vocabulary size ($v$). Compound features like $b(h/|mp|)$ represent derived quantities used to capture operation complexity.}
\label{tab:Feature Selection for Regressor} 
\begin{center}
\begin{tabular}{|l|r|}
\hline
\textbf{Operator} & \textbf{Workload Representation}   \\
\hline
Embedding       & $[bl, v/|mp|, d]$          \\
LayerNorm       & $[b, l, d]$          \\
RMSNorm         & $[b, l, d]$           \\
Linear1         & $[bl, d, 3d/|mp|]$                 \\
RoPE            & $[b, l, h/|mp|, d/h]$               \\
$QK^{\top}$     & $[b(h/|mp|), l, d/h, l]$    \\
Fillmask        & $[b, h/|mp|, l, d]$                     \\
Softmax         & $[b, h/|mp|, l, l]$                 \\
Fused Softmax   & $[b(h/|mp|), l, l]$                  \\
$\cdot V$       & $[b(h/|mp|), l, l, d/h]$       \\
Flash Attention & $[b, l, h/|mp|, d/h]$                      \\
Linear2         & $[bl, d/|mp|, d]$    \\
Linear3         & $[bl, d, 4d/|mp|]$    \\
Glue            & $[b, l, 4d/|mp|]$     \\
Linear4         & $[bl, 4d/|mp|, d]$    \\
Final\_Linear   & $[bl, d, v/|mp|]$ \\
Parallel Cross-entropy & $[b, l, v/|mp|]$ \\
MP\_All-reduce  & $[bld, |nodes|, |GPUs/nodes|]$ \\
DP\_All-reduce  & $[|entries|, |nodes|, |GPUs/node|]$ \\
DP\_All-gather  & $[|entries|, |nodes|, |GPUs/node|]$ \\
PP\_P2P         & $[bld/|mp|, |nodes|, |GPUs/node|]$ \\
Optimizer       & $[|mp|, dim, |encoders|]$    \\
\hline
\end{tabular}
\vspace{-2ex}
\end{center}
\end{table}

Specialized attention operators, such as $QK^{T}$ computation ($[b(h/|mp|), l, d/h, l]$) and value multiplication, i.e., $\cdot V$ ($[b(h/|mp|), l, l, d/h]$), define the computational complexity of self-attention mechanisms. Optimized implementations, such as Flash Attention ($[b, l, h/|mp|, d/h]$), reduce memory overhead by fusing multiple steps into a single kernel execution. The feature set further incorporates parallelized embedding layers, loss computation (parallel cross-entropy, $[b, l, v/|mp|]$), and optimizer overheads based on model configuration. 

Another critical category of operators is parallelism, which encompasses model parallel (MP), data parallel (DP), and pipeline parallel (PP) communication operations. These operators handle synchronization, gradient aggregation, and inter-stage data transfer in distributed training. For example, MP\_All-reduce ($[bld, |nodes|, |GPUs/nodes|]$) synchronizes activations or gradients across model-parallel partitions. Here, $bld$ represents the total volume of transferred data, $|nodes|$ accounts for inter-node synchronization overhead, and $|GPUs/nodes|$ captures intra-node communication efficiency, which depends on NVLink or PCIe topology. Other parallelism operators, such as DP\_All-reduce and PP\_P2P, similarly encode key workload and system parameters, adapting to their communication patterns.

System-level descriptors define GPU topology and hardware configurations, including interconnect types (NVLink, InfiniBand), GPU count per node, and architectural differences between hardware generations (e.g., Ampere vs. Hopper). The framework also differentiates pipeline stage roles (first, middle, last) to account for variations in activation propagation, workload distribution, and synchronization patterns across training steps.

\vspace{1ex}
\noindent \textbf{Feature Selection and Regression Considerations:}
The regressor input combines workload, parallelism, and system features for robust performance estimation. To avoid accuracy loss when extrapolating beyond benchmarks, we strategically sample high-impact configurations. Kernel fusion in modern frameworks can cause discrepancies between micro-benchmarks and real runtimes. We address this with framework-aware timing adjustments and iterative refinement based on real workload validation.

\subsection{Timeline Modeling of Distributed Training}

In large-scale LLM training, pipeline parallelism introduces unique performance considerations that our modeling framework must capture. This section details how we model pipeline execution, partitioning strategies, and their integration into our overall performance prediction methodology.

\vspace{1ex}
\noindent\textbf{Vocabulary Size Configuration and Dataset Characteristics:}
A critical implementation detail for vocabulary handling is the size alignment requirement. To optimize memory access patterns and computational efficiency, we ensure the vocabulary size is divisible by 128 times the number of model parallel partitions:

\begin{equation}
\text{divisibility\_factor} = 128 \times \text{num\_MP\_partitions}
\end{equation}

\vspace{-1em}
\begin{equation}
\text{vocab\_size} = \left\lceil \frac{\text{original\_vocab\_size}}{\text{divisibility\_factor}} \right\rceil \times \text{divisibility\_factor}
\end{equation}


For our training and evaluation, we used The Pile dataset~\cite{gao2020pile800gbdatasetdiverse}, a diverse 825GB English text corpus designed explicitly for language model pretraining. We processed this data using the GPT-NeoX-20B tokenizer with a vocabulary size of 50,257 tokens, providing realistic workloads for our performance modeling.

\vspace{1ex}
\noindent\textbf{Pipeline Partitioning Method:}
A key challenge in pipeline parallelism is model partitioning. Since two blocks (EmbeddingPipe and Pre-Transformer) precede the encoders and three blocks (Post-Transformer, NormPipe, and ParallelLinearPipe) follow them, encoder block allocation is computed using the following formulas to balance the workload across stages:
\begin{align}
\text{First stage encoders} &= \left\lceil\frac{\#\text{encoders} + 5}{\#\text{PP\_stages}}\right\rceil - 2 \\
\text{Middle stage encoders} &= \left\lfloor\frac{\#\text{encoders} + 5}{\#\text{PP\_stages}}\right\rfloor \\
\text{Last stage encoders} &= \left\lfloor\frac{\#\text{encoders} + 5}{\#\text{PP\_stages}}\right\rfloor - 3
\end{align}

\vspace{1ex}
\noindent\textbf{Number of Parameters for Pipeline Stages:}
We model the number of parameters per stage to predict the time cost of DP\_AllReduce and DP\_AllGather, as shown in Table~\ref{tab:PP Stage Parameters}. Based on Table~\ref{tab:Transformer Encoder Operators}, we use the following formula specifically for encoder parameters:

\begin{equation}
\#\text{encoder\_parameters} = 4d + \frac{8d(d+1)}{|mp|} + \frac{d(4d+1)}{|mp|}
\end{equation}

\begin{table}[htbp]
\small
\caption{Comprehensive breakdown of operators showing parameter shapes. Its dimensional requirements characterize each operator.}
\label{tab:Transformer Encoder Operators} 
\begin{center}
\begin{tabular}{|l|r|r|}
\hline
\textbf{Operator}  & \textbf{Parameters Shape}  \\
\hline
Parallel Embedding             & $[v/|mp|, d]  $      \\
\hline
LayerNorm                                 & $[d], [d]$        \\
Linear1                     & $[d, 3d/|mp|],[3d/|mp|]$           \\
Linear2                      & $[d/|mp|, d],[d]$   \\
LayerNorm                               & $[d], [d]$   \\
Linear3                         & $[d, 4d/|mp|],[4d/|mp|]$ \\
Linear4                 & $[4d/|mp|, d],[d]$ \\
\hline
LayerNorm                               & $[d], [d]$   \\
Final\_Linear           & $[d, v/|mp|]$ \\
\hline
\end{tabular}
\vspace{-2ex}
\end{center}
\end{table}




\begin{table}[htbp]
\caption{Distribution of parameters across pipeline parallel stages in a model parallel setting. The table quantifies parameter count for the first, middle, and last stages, where v represents vocabulary size, d hidden dimension, mp model parallel degree, and n the number of encoder layers per stage.}
\label{tab:PP Stage Parameters} 
\begin{center}
\begin{tabular}{|l|r|r|}
\hline
\textbf{PP Stage} & \textbf{\#Parameters} \\
\hline
First Stage       & $v\times d/|mp| + n\times\#encoder\_parameters $     \\
Middle Stages       & $n\times\#encoder\_parameters $   \\
Last Stage       & $n\times\#encoder\_parameters + 2d + v \times d/|mp|$     \\
\hline
\end{tabular}
\vspace{-2ex}
\end{center}
\end{table}



    
    

\begin{figure}[ht]
\centering
\includegraphics[width=0.4\textwidth]{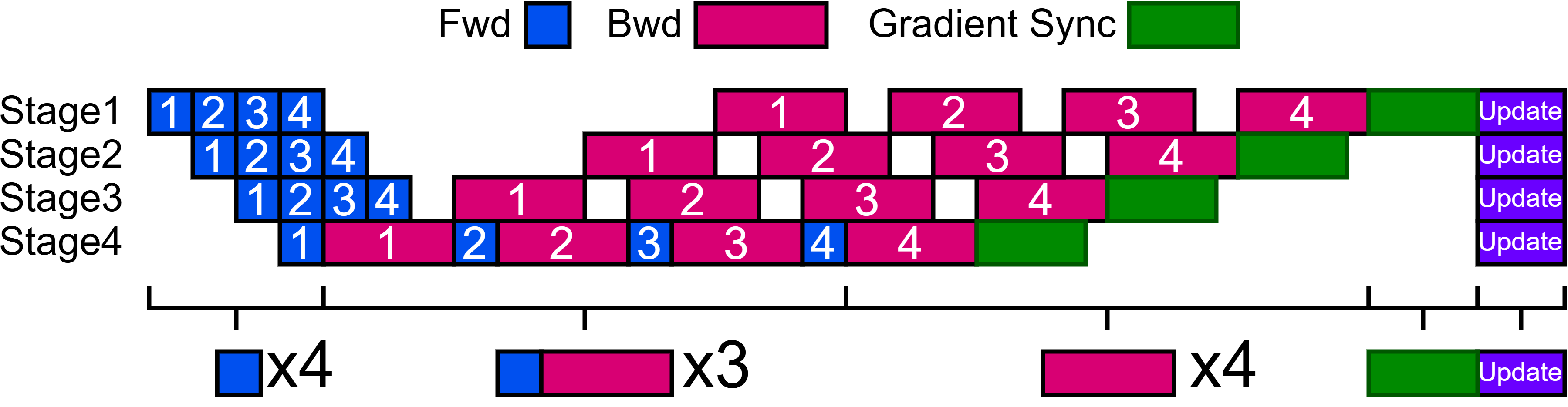}
\caption{Timeline of 1F1B Pipeline \& Data Parallelism. 1F1B pipeline parallelism refers to each pipeline stage executing the backward propagation in advance whenever possible. The gradient synchronization in data parallelism is hidden by the backward propagation of previous pipeline stages, except for stage 1.}
\label{fig:1F1B pipeline}
\end{figure} 

\vspace{1ex}
\noindent\textbf{Batch Runtime Calculation:}
To predict the complete training iteration time, we combine the pipeline execution with data-parallel synchronization and optimization steps. In Figure~\ref{fig:1F1B pipeline}, apart from the pipeline acceleration of forward and backward propagation, where faster stages are hidden, the gradient synchronization steps from the second-to-last stages are overlapped, and the update steps are also concealed by the stage with the maximum update time. With this consideration, the total time cost per parameter update for 1F1B pipeline and data parallelism is given by:
\vspace{2ex}
\begin{equation} \label{con:Timecost Per Update}
\begin{split}
    Runtime & = (\#Micro\_Batches - 1 + \#Pipeline\_Stages)  \\
             & \times (Max\_Fwd + Max\_Bwd) \\
             & + First\_Stage\_Gradient\_Synchronization \\
             & + Max\_Update
\end{split}
\end{equation}

Here $\#Micro\_Batches$ is the number of micro-batches that a mini-batch is divided into, and also represents the number of iterations performed before a parameter update occurs. $\#Pipeline\_Stages$ is the number of pipeline parallel partitions. Time cost of $PP\_P2P$ is assigned to the sender stages. $Max\_Fwd$ and $Max\_Bwd$ are the maximum forward and backward pass times across all pipeline stages. The time cost of $First\_Stage\_Gradient\_Synchronization$ step is equal to $DP\_Allreduce(\#Stage\_Parameters)$. $Max\_Update$ is the maximum of all stages' time cost sum of $Optimizer$ and $DP\_Allgather(\#Stage\_Parameters/|dp|)$. Time cost of $MP\_Allreduce$ in Parallel Cross-Entropy and Optimizer steps will be ignored due to the negligible communication volume.

\section{Results}

This section presents a comprehensive evaluation of our operator-level performance prediction methodology across diverse large-scale language models (LLMs) and hardware platforms.  We detail our experimental configurations, report component-level and end-to-end prediction accuracies, and provide insights into observed performance patterns across parallel strategies and architectural designs.

\begin{table}[htbp]
\small
\caption{Model configurations for the three target LLMs evaluated in this study. Each model's architecture is defined by its hidden dimension (d), sequence length (l), number of attention heads (h), and number of encoder layers. Additional configuration parameters include synchronization points and attention mechanism variants.}
\label{tab:LLM_config} 
\begin{center}
\begin{tabular}{|l|r|r|r|}
\hline
\textbf{Configs} & \textbf{GPT-20B} & \textbf{LLaMA-13B} & \textbf{Llemma-7B} \\
\hline
Hidden Dim(d)       & 6144  &   5120 &   4096\\
Sequence Length(l)      & 2048  &   2048 &   4096\\
Attention Heads(h)       & 64  &   40 &  32 \\
\#Encoders       & 44 &   40 &   32 \\
Encoder\_fwd Syncs      & 1 &   2 &  2\\
Encoder\_bwd Syncs      & 2 &   2 &  2\\
Fused Softmax      & True  &   True  &  False\\
Flash Attention       & False  &   False &   True\\
MLP Activation      & Gelu &   Gelu &  Gelu\\
Zero Stage      & 1 &   1 &  1\\
Optimizer      & FusedAdam &   FusedAdam &  FusedAdam\\
LayerNorm      & Basic &   RMS &  RMS\\
Precision      & FP16 &   FP16 &  FP16\\
Micro-Batch Size      & 4 &   4 &  4\\
Iters/Update     & 16 &   16 &  8\\
\hline 
\end{tabular}
\vspace{-4ex}
\end{center}
\end{table}

\subsection{Experimental Setup}

\noindent\textbf{Target Models and Testbeds:}
We evaluated GPT-20B, LLaMA-13B, and Llemma-7B (Table~\ref{tab:LLM_config}) on two distinct HPC platforms: Perlmutter (A100-SXM4) and Vista (H200-96GB HBM3) (Table~\ref{tab:spec}). These systems, spanning multi-GPU clusters and unified memory architectures, enable a comprehensive evaluation of our framework’s adaptability across diverse hardware paradigms.

\begin{table}[htbp]
    \caption{Cluster Specifications}
    \label{tab:spec}
    \footnotesize
    \centering
    \scalebox{1.0}{
    \begin{tabularx}{\columnwidth}{|l|X|X|}
        \hline
        \textbf{Specification} & \textbf{Perlmutter} & \textbf{TACC Vista} \\
        \hline
        \textbf{Processor} & AMD EPYC 7763 ("Milan") & NVIDIA Grace CPU Superchip \\
        \hline
        \textbf{Clock Speed} & 2.45 GHz & 3.1 GHz \\
        \hline
        \textbf{Cores} & 64 & 72 \\
        \hline
        \textbf{Memory} & 256 GB DDR4 & 120 GB LPDDR5 \\
        \hline
        \textbf{GPU} & NVIDIA A100-SXM4 & NVIDIA GH200 \\
        \hline
        \textbf{GPU Memory} & 40 GB HBM2 & 96 GB HBM3 \\
        \hline
        \textbf{GPUs/Node} & 4 & 1 \\
        \hline
        \textbf{Intra-Node Interconnect} & NVLink 3.0 (600 GB/s) & NVLink-C2C (900 GB/s) \\
        \hline
        \textbf{Inter-Node Interconnect} & Slingshot-10 (4 × 50 Gb/s NICs) & NVIDIA NDR InfiniBand (400 Gb/s) \\
        \hline
        \textbf{Scale} & Up to 32 Nodes (\textbf{128 A100}) & Up to 128 Nodes (\textbf{128 GH200})\\
        \hline
    \end{tabularx}
    }
\vspace{-1ex}
\end{table}

\begin{table}[htbp]
\caption{Parameter sampling ranges used for computing kernel benchmarks. The table shows the start value, step size/multiplier, and end value for each configuration parameter, including model parallelism (mp), batch size (b), attention heads (h), sequence length (l), and hidden dimension (d).}
\label{tab:config_range_comp} 
\begin{center}
\begin{tabular}{|l|r|r|r|}
\hline
\textbf{Configures} & \textbf{Start} & \textbf{Step} & \textbf{End} \\
\hline
$|mp|$         & 1   & $\times 2$    & 16    \\
$b$      & 4  & $\times 2$    & 8       \\
$h$     & 16     & +8    & 80       \\
$l$    & 1024     & +512    & 5120       \\
$d$     & 2048     & +512    & 8129       \\
\hline
\end{tabular}
\vspace{-2ex}
\end{center}
\end{table}

\begin{table}[htbp]
\small
\caption{Sampling ranges for communication kernel benchmarks. Each entry represents a vector of [\texttt{number of entries}, \texttt{number of processes}], covering different parallel communication patterns, including model parallel all-reduce, data parallel all-reduce/all-gather, and pipeline parallel point-to-point communication.}
\label{tab:config_range_commu} 

\begin{center}
\begin{tabular}{|l|r|r|r|}
\hline
\textbf{Configures} & \textbf{Start} & \textbf{Step} & \textbf{End} \\
\hline
MP\_AllReduce           & [2.09e7, 2]  & [+6.55e4, $\times2$]   & [1.34e8, 8]   \\
DP\_AllReduce      &[1.34e8, 2] & [+2.40e6, $\times2$]    & [1.20e9, 8]      \\
DP\_AllGather     & [1.34e8, 2]     & [+2.40e6, $\times2$]    & [6.01e8, 8]    \\
PP\_P2P    & [2.09e6, 2]     & [+6.55e4, 0]    & [1.34e8, 2]    \\
\hline
\end{tabular}
\vspace{-4ex}
\end{center}
\end{table}


\vspace{1ex}
\noindent\textbf{Benchmarking Methodology:} 
Our evaluation comprises micro-benchmarks on individual operators (Tables~\ref{tab:Feature Selection for Regressor}). Tables~\ref{tab:config_range_comp} and \ref{tab:config_range_commu} dictate the sampling spans for target LLMs in \ref{tab:LLM_config}.  
\subsection{Performance Stability}
In Table~\ref{tab:iteration_statistics}, Perlmutter shows high performance stability across all configurations, with less than 1\% variation between minimum and average training times. Vista, while faster due to its higher-end GPUs, exhibits greater variability (5.21\%–108.3\%) under different parallelism strategies. This instability stems from its single-GPU-per-node design, which forces all communication over the inter-node network—especially MP\_All-Reduce operations. In contrast, Perlmutter’s multi-GPU nodes enable intra-node pre-reduction, reducing inter-node communication. To mitigate variability, we use the minimum training batch cost as the prediction target.

\begin{table*}[htbp]
\centering
\caption{Statistics of training batch time cost measurement(second) of different model configurations and parallel strategies for both Perlmutter and TACC Vista platforms. Configuration notation (x-y-z) represents Pipeline-Model-Data parallelism degrees respectively.}
\label{tab:iteration_statistics} 
\begin{center}
\begin{tabular}{|l|c|c|c|c|c|c|c|c|c|c|}
\hline
\multirow{2}{*}{\textbf{Training Batch}} & \multicolumn{2}{c|}{\textbf{GPT-20B(4-4-8)}} & \multicolumn{2}{c|}{\textbf{GPT-20B(4-8-4)}} & \multicolumn{2}{c|}{\textbf{GPT-20B(8-4-4)}} & \multicolumn{2}{c|}{\textbf{LLaMA-13B(4-8-2)}} & \multicolumn{2}{c|}{\textbf{Llemma-7B(4-2-2)}} \\
\cline{2-11}
 & \textbf{P} & \textbf{V} & \textbf{P} & \textbf{V} & \textbf{P} & \textbf{V} & \textbf{P} & \textbf{V}  & \textbf{P} & \textbf{V}  \\
\hline
Minimum              &17.35    &7.13    & 42.53     &7.48  &9.23     & 6.39   & 48.37   & 6.76   & 9.57   & 5.04\\
\hline
Maximum              &17.56    &16.06    & 42.99     &21.23      &9.31    & 27.9    &49.08    & 9.40  &9.63   & 6.30\\
\hline
Average             & 17.43   &8.62   & 42.80     &12.75   & 9.27    &13.30    & 48.76     &7.54   & 9.59   &5.30 \\
\hline
\textbf{\% Increase of Average to Min} & \textbf{0.47}\%   &\textbf{20.98}\%   & \textbf{0.65}\%   &\textbf{70.38}\%  & \textbf{0.47}\%   & \textbf{108.30}\% & \textbf{0.80}\%    & \textbf{11.57}\%  & \textbf{0.16}\%   & \textbf{5.21}\%\\
\hline
\end{tabular}
\vspace{-2ex}
\end{center}
\end{table*}

\subsection{Prediction Errors and Analysis}

\label{sec:prediction-errors}

\begin{table*}[htbp]
\centering
\caption{Component-level prediction errors for operation timing based on measurements from the fastest training batch across various model scales and parallelization strategies. Results are shown for both Perlmutter and TACC Vista platforms. Configuration notation (x-y-z) represents Pipeline-Model-Data parallelism degrees respectively. The micro-batch size for all cases is 4, with the number of iterations for parameter updates set to 8 for \textbf{Llemma-7B(4-2-2)}, while all other cases use 16 iterations. The average relative errors of overall predictions are 4.98\% and 9.38\% for Perlmutter and Vista, respectively.}
\label{tab:module_Errors} 
\begin{center}
\begin{tabular}{|l|c|c|c|c|c|c|c|c|c|c|}
\hline
\multirow{2}{*}{\textbf{Component}} & \multicolumn{2}{c|}{\textbf{GPT-20B(4-4-8)}} & \multicolumn{2}{c|}{\textbf{GPT-20B(4-8-4)}} & \multicolumn{2}{c|}{\textbf{GPT-20B(8-4-4)}} & \multicolumn{2}{c|}{\textbf{LLaMA-13B(4-8-2)}} & \multicolumn{2}{c|}{\textbf{Llemma-7B(4-2-2)}} \\
\cline{2-11}
 & \textbf{P} & \textbf{V} & \textbf{P} & \textbf{V} & \textbf{P} & \textbf{V} & \textbf{P} & \textbf{V}  & \textbf{P} & \textbf{V}  \\
\hline
Encoder\_Fwd                &-13.60\%    &-11.47\%    & 1.47\%     &-14.17\%  &-12.89\%     & -14.37\%  & -1.84\%     & -10.43\%   & -2.52\%   & -2.24\%\\
\hline
Encoder\_Bwd                &-11.93\%    &-8.05\%    & 5.49\%     &-9.69\%  &-7.74\%      & -8.63\%    & -10.34\%    & -11.78\%  & -1.55\%   & -13.33\%\\
\hline
Stage\_Fwd\_Max             & -13.41\%   &-8.96\%    & 6.43\%     &-11.99\%  & -9.50\%    & -9.00\%    & 0.53\%     & -9.06\%   & -1.86\%   & 0.32\%\\
\hline
Stage\_Bwd\_Max             & -12.49\%   &-8.55\%    & 4.89\%      &-10.37\%  & -8.33\%    & -10.39\%   & -11.11\%    & -12.52\%  & -1.66\%   & -13.73\%\\
\hline
DP\_Allreduce(First\_stage)  & -1.90\%    &-49.69\%   & -21.51\%    &-8.43\%  & -2.06\%     & -6.36\%    & 5.06\%      & -14.96\%  & -31.21\%  & -12.02\%\\
\hline
DP\_Allgather(Max\_Update)  & -0.02\%    &-52.47\%   & -1.89\%     &-4.94\%  & 1.63\%      & -7.66\%    & -1.35\%     & 18.68\%   & 37.86\%   & 9.68\%\\
\hline
Max\_Update                 & 1.93\%     &-46.39\%   & -2.05\%     &-19.58\%  & -6.18\%     & -24.17\%   & 7.92\%      & 25.53\%   & -16.29\%    & 9.00\%\\
\hline
MP\_Allreduce               & -4.85\%   &2.33\%    & 10.67\%     &1.22\%  & -1.59\%        & -7.66\%    & 1.98\%     & 1.99\%   & 4.42\%    & 1.12\%\\
\hline
PP\_P2P                     & 3.16\%   &-32.07\%   & 3.84\%   &-34.14\%  & -1.65\%         & -33.06\%   & -7.64\%    & -36.26\%  & -21.98\%   & -3.82\%\\
\hline
\textbf{Overall}       & \textbf{-8.82}\%   &\textbf{-9.15}\%   & \textbf{3.94}\%   &\textbf{-15.16}\%  & \textbf{-5.87}\%   & \textbf{-8.41}\% & \textbf{-4.95}\%    & \textbf{-9.02}\%  & \textbf{1.30}\%   & \textbf{-5.18}\%\\
\hline
\end{tabular}
\vspace{-4ex}
\end{center}
\end{table*}



\begin{figure*}[ht]
\centering
\includegraphics[width=1\textwidth]{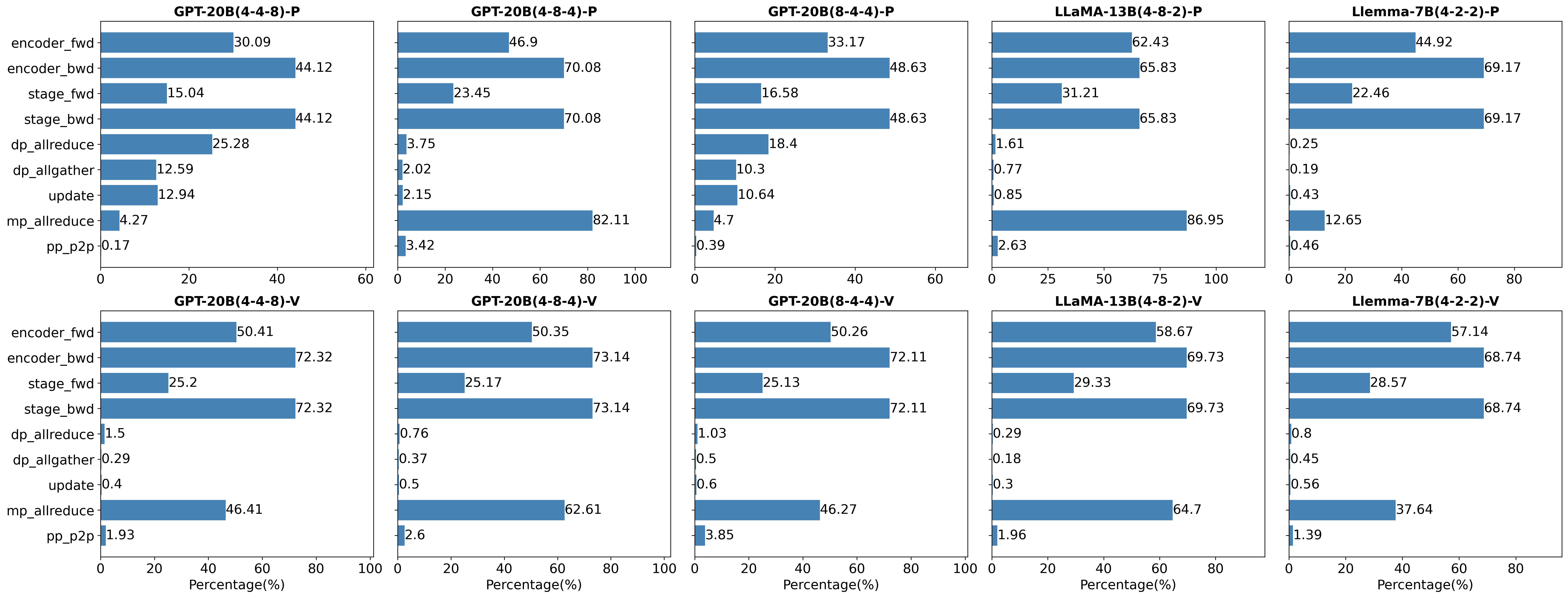}
\caption{Time cost proportions of different components across model configurations and platforms in our estimation for a single train batch. The top row shows Perlmutter results while the bottom row shows Vista results. The percentages represent time proportions relative to total runtime. Thus, they don't add up to 100\%, since not all components are independently of each other. } 
\label{fig:Component Proportion}
\vspace{-2ex}
\end{figure*}

We evaluate the accuracy of our performance model across both component-level operations and end-to-end training iterations, leveraging diverse model configurations and parallelization strategies. Table~\ref{tab:module_Errors} quantifies the per-component prediction errors and reports minimum end-to-end errors across different HPC platforms. Figure~\ref{fig:Component Proportion} provides context by illustrating the estimated runtime breakdown across operations. Note that summing all component percentages yields more than 100\% of the runtime. This is because not all components execute in isolation or strictly sequentially; only the major independent phases (Stage\_Fwd, Stage\_Bwd, DP\_Allreduce, and the Update step) are mutually exclusive, and their proportions sum to the total runtime.

\vspace{1ex} \noindent \textbf{Component-Level Accuracy:}  
The majority of runtime is dominated by computational operations, particularly encoder forward or backward passes and pipeline stage execution, which account for 70–95\% of total runtime as evidenced by Figure~\ref{fig:Component Proportion}. These compute-heavy components are predicted with high accuracy, with sub-3\% errors for smaller models (e.g., LLemma-7B on Perlmutter) and 10–15\% for large models such as GPT-20B.  

While communication operations such as DP\_Allreduce, DP\_Allgather, and PP\_P2P exhibit higher prediction errors (Table~\ref{tab:module_Errors})—occasionally exceeding 50\%—this discrepancy has minimal impact on overall model accuracy. These components account for less than 5\% of the total iteration time across all configurations (Figure~\ref{fig:Component Proportion}), and their inaccuracies are amortized within the broader prediction pipeline. Crucially, our modeling framework prioritizes high-accuracy regressors for the dominant contributors to runtime (e.g., encoder forward/backward passes and pipeline stages), which collectively drive end-to-end performance. The elevated error rates for certain communication operations are, therefore, a benign artifact of our design choice to allocate modeling capacity proportionally to component impact. Moreover, our system-level timeline model accounts for overlap and concurrency, mitigating the effect of local communication inaccuracies on final predictions. 


On the other hand, in communication-dominated scenarios, such as those involving MP\_Allreduce, which constitutes a substantial portion of the total runtime across all models on TACC Vista, as well as for configurations with eight-way model parallelism on Perlmutter. This is expected, as MP\_Allreduce is invoked once or twice during each encoder forward pass and twice during each backward pass, and always entails inter-node communication in these settings. Despite its prominence, the prediction error for MP\_Allreduce remains consistently low, staying below 5\% in most cases and peaking at 10.67\%. This robustness is attributed to its high invocation frequency, which statistically dampens the impact of transient synchronization overheads and network variability, thereby ensuring stable and accurate modeling even under communication-intensive workloads.

\vspace{1ex} \noindent \textbf{End-to-End Prediction Evaluation:}  
The overall runtime function~\ref{con:Timecost Per Update} provides an estimate with communication errors minimized through a heuristic, pipeline-aware cost model that aligns with the overlap capabilities demonstrated in state-of-the-art pipeline parallelism studies. The low prediction errors reflect the overall robustness of the function, despite localized inaccuracies—particularly those arising from the inherently stochastic nature of real-world communication. 

As summarized in Table~\ref{tab:module_Errors}, our framework demonstrates promising overall accuracy, with Perlmutter maintaining a balanced error range (-8.82\% to 3.94\%) and Vista exhibiting a consistent underestimation trend (-5.18\% to -15.16\%). On Vista, the -15.16\% error for GPT-20B(4-8-4) is primarily due to the failure to capture its actual minimum training time, which should be lower than that of GPT-20B(4-4-8), given that the latter processes twice the effective batch size. The mean end-to-end prediction error is 4.98\% on Perlmutter and 9.38\% on Vista, indicating that our model offers a \textit{conservative upper-bound estimate} for GH200-based architectures. This is considered a good sign and helps reduce the risk of under-provisioning compute resources. 

Prediction accuracy correlates with parallelization strategy complexity. On Perlmutter, high data-parallel configurations (4-4-8) show a -8.8\% error, while more balanced model-parallel configurations (4-8-4) achieve a low 3.9\% error, indicating that moderate model parallelism yields the highest prediction fidelity. Increasing pipeline depth (8-4-4) results in a -5.9\% error, reinforcing that higher pipeline parallelism introduces additional variability. These trends suggest that data-parallel-heavy and deep pipeline-parallel configurations pose more significant modeling challenges, while balanced parallelism setups are more straightforward to predict with high fidelity.

Smaller-scale models exhibit the highest accuracy. The Llemma-7B runs (16 GPUs) achieve 1.3\% and -5.18\% error on Perlmutter and Vista, respectively, which closely match the actual runtime on Perlmutter and show a slightly conservative estimate on Vista. Errors increase moderately as the model scales. However, even the largest 128-GPU GPT-20B jobs remain within a practical 5–15\% range, demonstrating that our framework effectively captures compute and communication patterns across different scales and parallelization strategies. This level of accuracy is particularly valuable for HPC scheduling and large-scale deep learning deployments, where even minor performance deviations can lead to substantial resource inefficiencies.

\vspace{1ex} \noindent \textbf{Architectural Impact on Prediction Accuracy:}  
Perlmutter’s multi-GPU nodes (A100-SXM4 with NVLink3) allow some collectives to run locally, improving prediction stability. In contrast, Vista’s single-GPU-per-node GH200 design relies entirely on InfiniBand, making it more vulnerable to network jitter and contention. The higher errors on Vista(Table~\ref{tab:module_Errors}) likely stem from these unpredictable network factors rather than modeling inaccuracies. Large-scale runs (up to 128 nodes) further amplify such variability, as our model remains accurate for computation estimation but sensitive to inter-node communication noise.

\vspace{1ex} \noindent \textbf{Implications for HPC Resource Allocation:}  
Overall, our modeling methodology consistently achieves a 5–10\% error margin in end-to-end runtime estimation, providing actionable guidance for HPC scheduling and resource allocation. By accurately predicting the runtime of LLM and distributed training configurations, the framework enables informed parallel strategy tuning and targeted optimization for either resource utilization or time-to-solution in large-scale training scenarios.

\section{Related Work}
\noindent \textbf{Computation and Communication Modeling Approaches:}
Early modeling techniques such as LogP~\cite{culler1993logp} and BSP~\cite{valiant1990bridging} characterized distributed computation using parameters such as latency, bandwidth, and synchronization overheads, and have been applied to supercomputing~\cite{Dongarra1992}, parallel programming~\cite{Zheng2005, Balasundaram1991}, and MPI communication~\cite{Chan2007, Thakur2005}. While foundational, these models often require estimating abundant parameters. For instance, LogP and its variants involve at least four parameters, and extensions like LogGOPS~\cite{hoefler2010loggopsim} introduce over seven, just for inter-node communication alone, making them increasingly difficult to calibrate for modern heterogeneous systems. For GPU kernel performance, analytical models~\cite{hong2009analytical, baghsorkhi2010adaptive, zhang2011quantitative} offer architecture-specific insights but rely on detailed instruction-level profiling and microarchitectural analysis, limiting practicality at scale. In contrast, our approach combines lightweight sampling with tree-based regressors, significantly reducing modeling complexity while maintaining high accuracy across diverse hardware and training configurations.

Recent works, including parallel strategy search in Alpa ~\cite{280874} and a subsequent study~\cite{yun2023fast} rely on comparisons based on implicit partial workload estimations. A systematic analytical method \cite{10763669} is closely related to ours in the context of training time estimation. Their results show prediction errors of less than 10\% for Nvidia’s data~\cite{korthikanti2023reducing, narayanan2021efficient} across a wide range of configurations, from 8 to 3072 GPUs and LLM sizes ranging from 22B to 1008B. However, their implementation are not openly available, impeding direct comparisons.

\vspace{1ex}
\noindent \textbf{Bottom-up and Top-down Modeling Approaches: }
Bottom-up approaches model GPU performance by analyzing low-level hardware characteristics. The IPP model~\cite{hong2010integrated} pioneered analytical GPU modeling with 8.9\% error, while Boyer et al.~\cite{boyer2013improving} extended this to include data transfers (8\% error). For deep learning, DeLTA~\cite{lym2019delta} achieved 7.9\% error by modeling arithmetic intensity and memory traffic, though requiring low-level profiling not feasible without hardware access. GROPHECY~\cite{meng2011grophecy} predicted GPU performance based on CPU code characteristics with 17\% error.

Top-down approaches analyze application-level behavior without detailed hardware knowledge. Baldini et al.~\cite{baldini2014predicting} used supervised learning for GPU performance prediction (23\% error), while Lee et al.'s GCoM~\cite{lee2022gcom} modeled GPU communication with 25.5\% error. Salaria et al.~\cite{salaria2019learning} improved accuracy to 9.4\% by incorporating architectural information.


Our framework uses a hybrid bottom-up approach based on operator-level decomposition and hardware-aware regressors. It accurately models compute and communication behavior using microbenchmark-guided sampling, achieving high accuracy and generalization across hardware platforms without instruction-level profiling or full end-to-end sampling.

\vspace{1ex}
\noindent \textbf{DNN-Specific Modeling:}
Specialized approaches address neural networks' unique computational patterns. Zancato et al.~\cite{zancato2020predictingtrainingtimetraining} predicted training convergence time without actual training (20\% error). Frameworks like DayDream~\cite{zhu2020daydream} model training with dependency graphs, while DTS~\cite{esposito2022dts} estimates time across distributed learning paradigms, though both require substantial profiling. Pinel et al.~\cite{pinel2020evolving} explored co-evolutionary approaches but with limited scalability to modern distributed settings.

\section{Conclusion}

We presented an operator-level performance prediction framework for LLM training that achieves average prediction errors of 4.98\% on Perlmutter and 9.38\% on Vista across diverse model configurations. By decomposing complex architectures into fundamental primitives and using specialized regression models, our approach accurately forecasts training performance without requiring costly experimentation. Communication operations remain the most challenging to predict precisely, particularly PP\_P2P operations on unified memory architectures. This suggests opportunities for improved network modeling in future work. Additional research directions include integration with job scheduling systems, adaptation to emerging hardware architectures, and incorporating energy efficiency metrics.
As LLMs continue to scale, accurate performance prediction becomes increasingly critical for resource optimization. Our framework’s strong generalization across diverse models, parallelization strategies, and computing clusters enables informed decisions on resource allocation, promoting more efficient utilization of scarce HPC resources.


\bibliographystyle{./IEEEtran}
\bibliography{./IEEEabrv,./Reference}

\end{document}